\documentclass[onecollarge]{svjour2}       
\usepackage{graphicx}
\def\geqsim{\mathbin{\;\raise1pt\hbox{$>$}\kern-8pt\lower3pt\hbox{$\sim$}\;}}
\def\leqsim{\mathbin{\;\raise1pt\hbox{$<$}\kern-8pt\lower3pt\hbox{$\sim$}\;}}

\begin{document}

\title{Nonlinearly charged dilatonic black holes and their
Brans-Dicke counterpart: Energy dependent spacetime}
\author{S. H. Hendi$^{1,2}$\thanks{\emph{Present address:} hendi@shirazu.ac.ir} and M. S. Talezadeh$^{1}$} \institute{$^1$Physics
Department and Biruni Observatory, College of Sciences, Shiraz
University, Shiraz 71454, Iran\\
$^2$Research Institute for Astrophysics and Astronomy of Maragha
(RIAAM), P.O. Box 55134-441, Maragha, Iran}

\date{Received:  / Accepted: }

\maketitle

\begin{abstract}
Regarding the wide applications of dilaton gravity in the presence
of electrodynamics, we introduce a suitable Lagrangian for the
coupling of dilaton with gauge field. There are various
Lagrangians which show the coupling between scalar fields and
electrodynamics with correct special situations. In this paper,
taking into account conformal transformation of Brans-Dicke theory
with an electrodynamics Lagrangian, we show that how scalar field
should couple with electrodynamics in dilaton gravity. In other
words, in order to introduce a correct Lagrangian of dilaton
gravity, one should check at least two requirements: compatibility
with Brans-Dicke theory and appropriate special situations.
Finally, we apply the mentioned method to obtain analytical
solutions of dilaton-Born-Infeld and Brans-Dicke-Born-Infeld
theories with energy dependent spacetime.
\end{abstract}

\section{Introduction\label{Intro}}

Einstein gravity of general relativity is one of the successful
theories of $20^{th}$ century. Although this theory can explain
the dynamics of our solar system well enough, it cannot describe
the high curvature regimes. In other words, one may use its
generalization for describing the particle motion near the black
hole horizon. Another problem of Einstein gravity is its
inaccurate result for the accelerated expansion of the universe.
In addition, Einstein gravity does not accommodate either Mach's
principle or Dirac's large-number hypothesis. Between various
modifications of Einstein gravity, Brans and Dicke were pioneers
in studying the alternative theories known as Brans-Dicke (BD)
theory \cite{BD}. BD theory is a straightforward way to
disentangle further degrees of freedom which are not present in
the usual Hilbert-Einstein action. Regarding the four dimensional
stationary BD solution, one finds it is just the Kerr solution
with a constant scalar field \cite{Gao4}. Beside the vacuum
solutions, Cai and Myung showed that the $4-$dimensional charged
black hole solution of BD theory is just the
Reissner--Nordstr\"{o}m (RN) solution with a trivial scalar field
\cite{sh51,sh52,sh551,sh552}. This is because the stress energy
tensor of Maxwell field is traceless only in $4-$dimension, and
therefore, the action of Maxwell field is not invariant under the
conformal transformation in higher dimensions. In order to find
the differences between BD theory and Einstein gravity, one may
regard higher dimensions. As an example it was found that higher
dimensional charged black holes in BD theory would be the RN
solution with a non-trivial scalar field. Classical and quantum
aspects black holes and gravitational collapse in Brans-Dicke (BD)
theory have been investigated in literature
\cite{Gao31,Gao32,Gao33,Gao34}.

Another distinguishing mark of BD theory is the existence of a
conformal mapping between this theory and Einstein gravity. A
conformal mapping is a diffeomorphism between two metrizable
bundles if the pulled back bundle is conformally equivalent to the
original one. The conformal transformation technique is one of the
strong and useful mathematical tools for investigating the
scalar-tensor theories of gravity \cite{CTa,CTb,CTc}. In other
words, using conformal transformation, one can represent gravity
in two conformally related frames: the Jordan or string frame (
scalar field is non-minimally coupled to the metric tensor) and
the Einstein frame (scalar field is minimally coupled to the
metric tensor) \cite{BDvsDilaton1,BDvsDilaton2,BDvsDilaton3}.
Investigation of these frames enables one to explore the relation
between different physical theories and hence generate the new
exact solutions \cite{CT2a,CT2b}. In addition, it was shown that
the higher-order theories of gravity can be related to the
non-minimally coupled version of scalar tensor gravity
\cite{Capoz}. In general, equipped with the conformal
transformations enables one to investigate gravitational theories
with more details since one can exchange, via these
transformations, a model of gravity to another where the relations
become more easier to manipulate. Although there are many studies
related to conformal transformation between the Einstein and
Jordan frames, there was not a user-friendly rule for presenting
mathematical properties of this transformation. Furthermore, the
standard coupling between dilaton and gauge fields has not been
explored with a systematic details. These are our main motivations
to present conformal transformation for obtaining standard
gauge-gravity Lagrangians with a methodical way. In this work, we
obtain a new criterion for conformal consistency of dilaton
gravity coupled with electrodynamics.

The main goal of the present work is finding how an arbitrary
Lagrangian of electrodynamics can couple with dilaton field in
Einstein frame. In this regard, we show that most of previous
introduced Lagrangians are not consistent with any BD counterpart.
In order to obtain original solutions, we focus on the challenging
Lagrangian, dilatonic Born-Infeld theory, in an energy dependent
spacetime, so-called gravity's rainbow. The motivation of
considering gravity's rainbow comes from Horava-Lifshitz gravity
\cite{HoravaPRD,HoravaPRL}, in which help us to regard different
scaling of space and time for type IIA and IIB string theory
\cite{Gregory2010,Burda2014}, AdS/CFT correspondence
\cite{Gubser2009,Ong2011,Kachru2014} and dilaton black objects
\cite{Goldstein2010,Bertoldi2010,HendidilatonicRb,Tarrio2011}. The
relation between the Horava-Lifshitz gravity and gravity's rainbow
was investigated in Ref. \cite{Garattini2015}. Gravity's rainbow
is based on the modification of usual energy-momentum dispersion
relation in the UV limit. It is worthwhile to mention that such
modification has also been regarded in different subjects, such
as: discrete spacetime \cite{Hooft}, spin-network in loop quantum
gravity (LQG) \cite{Gambini}, spacetime foam \cite{Camelia},
non-commutative geometry \cite{Carroll,FaizalMPLA} and ghost
condensation \cite{FaizalJPA}.

Motivated by the recent results mentioned above, we obtain the
correct Lagrangian of gauge-dilaton coupling with corresponding
field equations. We also introduce the correct action and field
equations of some interesting models of nonlinear electrodynamics
in dilaton gravity. Finally, we consider an energy dependent
spacetime to obtain black hole solutions of BD-BI and dilatonic BI
theories.

\section{Field equations and conformal transformations \label{FE}}

The action of $(n+1)$- dimensional BD theory with a scalar field $\Phi $ and
a self-interacting potential $V(\Phi )$ in the presence of a matter field
can be written as
\begin{equation}
I=-\frac{1}{16\pi }\int d^{n+1}x\sqrt{-g}\left[ \Phi \mathcal{R}+\mathcal{L}%
(\Phi ,\nabla \Phi )+\mathcal{L}_{m}(\Psi )\right] ,  \label{IBD}
\end{equation}%
in which $\mathcal{L}(\Phi ,\nabla \Phi )$ is an arbitrary Lagrangian of
scalar field and its first derivative and $\mathcal{L}_{m}$ is Lagrangian of
pure matter field with field strength $\Psi $. The standard choice for the
Lagrangian of scalar field is
\begin{equation}
\mathcal{L}(\Phi ,\nabla \Phi )=-\frac{\omega }{\Phi }(\nabla \Phi
)^{2}-V(\Phi ),  \label{LPHI}
\end{equation}%
where the first and second expressions are, respectively,
characteristic as kinetic and potential terms in string frame.
Now, we would like to obtain electrodynamic Lagrangian coupled
with scalar field in Einstein frame. Indeed, via the conformal
transformation \cite{sh51,sh52,sh551,sh552} the BD theory can be
transformed into the Einstein-dilaton gravity with a minimally
coupled scalar dilaton field. We should note that the suitable
conformal transformation comes from the fact that $\sqrt{-g}\Phi
\mathcal{R}$ term in Jordan frame should transform to
$\sqrt{-\bar{g}}\bar{\mathcal{R}}$ in
Einstein frame. Suitable conformal transformation can be written as%
\begin{eqnarray}
\bar{g}_{\mu \nu } &=&\Phi ^{2/(n-1)}g_{\mu \nu },  \nonumber \\
\bar{\Phi} &=&\frac{n-3}{4\alpha }\ln \Phi ,  \label{CTeq}
\end{eqnarray}%
where
\begin{equation}
\alpha =\frac{(n-3)[(n-1)\omega +n]^{-1/2}}{2}.
\end{equation}%
It was shown that all functions and quantities in Jordan frame (${g}_{\mu
\nu }$, ${\Phi }$ and $\Psi $) can be transformed into Einstein frame ($\bar{%
g}_{\mu \nu }$, $\bar{\Phi}$ and $\bar{\Psi}$). Applying the mentioned
conformal transformation on the BD action (\ref{IBD}), one can obtain
corresponding action in Einstein frame
\begin{equation}
\bar{I}=-\frac{1}{16\pi }\int_{\mathcal{M}}d^{n+1}x\sqrt{-\bar{g}}\left\{
\bar{\mathcal{R}}+\bar{\mathcal{L}}(\bar{\Phi},\bar{\nabla}\bar{\Phi})+\bar{%
\mathcal{L}}(\bar{\Psi},\bar{\Phi})\right\} ,  \label{IEin}
\end{equation}%
where $\bar{\mathcal{R}}$ and $\bar{\nabla}$ are, respectively, the Ricci
scalar and covariant derivative corresponding to the metric $\bar{g}_{\mu
\nu }$, and $\bar{\mathcal{L}}(\bar{\Phi},\bar{\nabla}\bar{\Phi})$ is
(transformed) Lagrangian of scalar field and $\bar{\mathcal{L}}(\bar{\Psi},%
\bar{\Phi})$ is Lagrangian of matter field which is coupled with
scalar field after transformation. Considering Eq. (\ref{LPHI}),
one obtains
\begin{equation}
\bar{\mathcal{L}}({\bar{\Phi}},\bar{\nabla}{\bar{\Phi}})=-\frac{4}{n-1}(\bar{%
\nabla}\bar{\Phi})^{2}-\bar{V}(\bar{\Phi}),
\end{equation}%
where the first and second expressions are, respectively, corresponding to
the kinetic and potential terms in Einstein frame. It is easy to find $\bar{V%
}(\bar{\Phi})$ is
\begin{equation}
\bar{V}(\bar{\Phi})=\Phi ^{-(n+1)/(n-1)}V(\Phi ).  \label{PhiPhi}
\end{equation}

Now, we are in a position to choose a special matter field. We consider
general form Lagrangian of electrodynamics as a matter field. Thus, we can
write
\begin{equation}
\mathcal{L}_{m}(\Psi )=\mathcal{L}(\mathcal{F}),  \label{LF}
\end{equation}
where $\mathcal{L}(\mathcal{F})$ is an arbitrary Lagrangian of
electrodynamics in which $\mathcal{F}=F_{ab}F^{ab}$ is the Maxwell
invariant and $F_{ab}=\partial _{a}A_{b}-\partial _{b}A_{a}$ is
the electromagnetic field tensor with the gauge potential $A_{a}$.
For various explicit forms of (linear and nonlinear)
$\mathcal{L}(\mathcal{F})$, we refer the reader to
\cite{HendiBlackString,HendiJHEP1,HendiJHEP2}.

The total Lagrangian of BD theory in the presence of electromagnetic field
is
\begin{equation}
I_{G}=-\frac{1}{16\pi }\int_{\mathcal{M}}d^{n+1}x\sqrt{-g}\left( \Phi
\mathcal{R}-\frac{\omega }{\Phi }(\nabla \Phi )^{2}-V(\Phi )+%
\mathcal{L}(\mathcal{F})\right) ,  \label{acBD}
\end{equation}%
where we can obtain its corresponding equations of motion with the following
explicit forms
\begin{eqnarray}
G_{\mu \nu } &=&\frac{\omega }{\Phi ^{2}}\left( \nabla _{\mu }\Phi \nabla
_{\nu }\Phi -\frac{1}{2}g_{\mu \nu }(\nabla \Phi )^{2}\right) -\frac{V(\Phi )%
}{2\Phi }g_{\mu \nu }+\frac{1}{\Phi }\left( \nabla _{\mu }\nabla _{\nu }\Phi
-g_{\mu \nu }\nabla ^{2}\Phi \right)  \nonumber \\
&&+\frac{2}{\Phi }\left( \frac{1}{4}g_{\mu \nu }\mathcal{L}(\mathcal{F}%
)-F_{\mu \lambda }F_{\nu }^{\;\lambda }\mathcal{L}_{\mathcal{F}}\right) ,
\label{field01}
\end{eqnarray}%
\begin{equation}
\nabla ^{2}\Phi =\frac{\left( n+1\right) \mathcal{L}(\mathcal{F})-4\mathcal{%
FL}_{\mathcal{F}}}{2\left[ \left( n-1\right) \omega +n\right] }+\frac{1}{2%
\left[ \left( n-1\right) \omega +n)\right] }\left[ (n-1)\Phi \frac{dV(\Phi )%
}{d\Phi }-\left( n+1\right) V(\Phi )\right] ,  \label{field02}
\end{equation}%
\begin{equation}
\nabla _{\mu }\left( \mathcal{L}_{\mathcal{F}}F^{\mu \nu }\right) =0,
\label{field03}
\end{equation}%
where $G_{\mu \nu }$ and $\nabla _{\mu }$ are, respectively, the Einstein
tensor and covariant derivative of manifold $\mathcal{M}$ with metric $%
g_{\mu \nu }$ and $\mathcal{L}_{\mathcal{F}}=\frac{d\mathcal{L}(\mathcal{F})%
}{d\mathcal{F}}$. Now, we would like to obtain electrodynamic
Lagrangian coupled with scalar field in Einstein frame. Indeed,
via the conformal transformation \cite{sh51,sh52,sh551,sh552} the
BD theory can be transformed into the Einstein-dilaton gravity
with a minimally coupled scalar dilaton field. Applying the
mentioned conformal transformation, we obtain
\begin{equation}
\bar{I}_{G}=-\frac{1}{16\pi }\int_{\mathcal{M}}d^{n+1}x\sqrt{-\bar{g}}%
\left\{ \bar{\mathcal{R}}-\frac{4}{n-1}(\bar{\nabla}\bar{\Phi})^{2}-\bar{V}(%
\bar{\Phi})+\bar{\mathcal{L}}(\bar{\mathcal{F}},\bar{\Phi})\right\} .
\label{con-ac}
\end{equation}

One can obtain the equations of motion by varying this action
(\ref{con-ac}) with respect to $\bar{g}_{\mu \nu }$, $\bar{\Phi}$
and $\bar{F}_{\mu \nu }$
\begin{eqnarray}
\bar{\mathcal{R}}_{\mu \nu } &=&\frac{4}{n-1}\left( \bar{\nabla}_{\mu }\bar{%
\Phi}\bar{\nabla}_{\nu }\bar{\Phi}+\frac{1}{4}\bar{V}\bar{g}_{\mu \nu
}\right) -2Ae^{m\alpha \bar{\Phi}}Be^{k\alpha \bar{\Phi}}\bar{\mathcal{L}}_{%
\bar{Y}}\bar{F}_{\mu \eta }\bar{F}_{\nu }^{\eta }  \nonumber \\
&&+\frac{A}{n-1}e^{m\alpha \bar{\Phi}}\left[ 2\bar{Y}\bar{\mathcal{L}}_{\bar{%
Y}}-\bar{\mathcal{L}}(\bar{Y})\right] \bar{g}_{\mu \nu },  \label{fieldc1}
\end{eqnarray}
\begin{equation}
\bar{\nabla}^{2}\bar{\Phi}=\frac{n-1}{8}\frac{\partial \bar{V}}{\partial
\bar{\Phi}}-A\alpha (\frac{n-1}{8})e^{m\alpha \bar{\Phi}}\left[ m\bar{%
\mathcal{L}}(\bar{Y})+k\bar{Y}\bar{\mathcal{L}}_{\bar{Y}}\right] ,
\label{fieldc2}
\end{equation}%
\begin{equation}
\bar{\nabla}_{\mu }\left( e^{(m+k)\alpha \bar{\Phi}}\bar{\mathcal{L}}_{\bar{Y%
}}\bar{F}^{\mu \nu }\right) =0  \label{fieldc3}
\end{equation}%
where $\bar{\mathcal{L}}_{\bar{Y}}=\frac{\partial \bar{\mathcal{L}}(\bar{Y})%
}{\partial \bar{Y}}$ and we use the following familiar notation
\begin{eqnarray}
\bar{L}(\bar{\mathcal{F}},\bar{\Phi}) &=&Ae^{m\alpha \bar{\Phi}}\bar{L}(\bar{%
Y}),  \label{L(FP)} \\
\bar{Y} &=&Be^{k\alpha \bar{\Phi}}\bar{\mathcal{F}},  \label{Y}
\end{eqnarray}%
in which $\alpha$ is an arbitrary constant that governs the
strength between the dilaton and electromagnetic fields, and $A$
and $B$ are two constants depending on the theory of
electrodynamics.
It is worthwhile to mention that in order to have a consistent $\bar{L}(%
\bar{\mathcal{F}},\bar{\Phi})$ with that of BD theory, we obtain
\begin{eqnarray*}
\Phi ^{-4/(n-1)}e^{k\alpha \overline{\Phi }} &=&1, \\
\Phi ^{(n+1)/(n-1)}e^{m\alpha \overline{\Phi }} &=&1.
\end{eqnarray*}%
Using Eq. (\ref{CTeq}), one finds
\begin{eqnarray}
k &=&\frac{16}{(n-1)(n-3)},  \label{k} \\
m &=&-\frac{4(n+1)}{(n-1)(n-3)}.  \label{m}
\end{eqnarray}%
Taking into account the mentioned calculations, we find that considering an
arbitrary nonlinear electrodynamics model in Jordan frame (BD theory) leads
to a corresponding $\bar{L}(\bar{\mathcal{F}},\bar{\Phi})$ in Einstein frame
(dilaton gravity) with the following form%
\begin{eqnarray}
\bar{L}(\bar{\mathcal{F}},\bar{\Phi}) &=&Ae^{-\frac{4(n+1)\alpha \bar{\Phi}}{%
(n-1)(n-3)}}\bar{L}(\bar{Y}),  \label{LFP2} \\
\bar{Y} &=&Be^{\frac{16\alpha \bar{\Phi}}{(n-1)(n-3)}}\bar{\mathcal{F}}.
\label{Y2}
\end{eqnarray}

\section{Case Studies}
In this section, we consider BD action with various models of
linear and nonlinear electrodynamics and try to obtain the correct
Lagrangian of gauge-dilaton coupling in the Einstein frame by
using the mentioned prescription.

\subsection{Linear Maxwell source:}
Maxwell electrodynamics is the most familiar theory between other
abelian gauge fields. It is a linear theory with excellent
consequences in the classical physics. Thus, as the first example,
we consider the Lagrangian of linear Maxwell field
\begin{equation}
\mathcal{L}(\mathcal{F})=-\mathcal{F},
\end{equation}%
and therefore $\bar{L}(\bar{Y})=\bar{Y}$ with $A=1$ and $B=-1$. Taking into
account Eqs. (\ref{LFP2}) and (\ref{Y2}), we find%
\begin{equation}
\bar{L}(\bar{\mathcal{F}},\bar{\Phi})=-e^{-\frac{4(n+1)\alpha \bar{\Phi}}{%
(n-1)(n-3)}}e^{\frac{16\alpha \bar{\Phi}}{(n-1)(n-3)}}\bar{\mathcal{F}}=-e^{%
\frac{-4\alpha \bar{\Phi}}{(n-1)}}\bar{\mathcal{F}},  \label{LFPMaxwell}
\end{equation}%
which is coupled Lagrangian of dilaton-Maxwell theory in which
introduced in Ref. \cite{sh51,sh52}.

\subsection{Born-Infeld theory}

Nonlinear electromagnetic theories have been considered in the
context of superstring theory. In other words, it was shown that
the Born-Infeld (BI) type Lagrangian may be regarded as all order
loop corrections to gravity \cite{BIPapers1,BIPapers2}. In
addition, one may find that the dynamics of D-branes are related
to the BI action \cite{Dbrane}. Considering BI type
electrodynamics coupled to the gravitational fields leads to black
hole solutions with interesting properties. Here, we apply the
obtained consequences to the case of Born-Infeld theory
\cite{BIa,BIb,BIPapers3,BIPapers4,BIPapers5,BIPapers6}
\begin{equation}
\mathcal{L}(\mathcal{F})=4\beta ^{2}\left( 1-\sqrt{1+\frac{\mathcal{F}}{%
2\beta ^{2}}}\right).  \label{LBI}
\end{equation}

Considering Eqs. (\ref{LFP2}) and (\ref{Y2}), and using
$\bar{L}(\bar{Y})=1-\sqrt{1+\bar{Y}}$ with $A=4\beta ^{2}$ and
$B=\frac{1}{2\beta ^{2}}$, we obtain the following lagrangian for
the coupling Lagrangian of dilaton-Born-Infeld theory
\begin{equation}
\bar{L}(\bar{\mathcal{F}},\bar{\Phi})=4\beta ^{2}e^{-\frac{4(n+1)\alpha \bar{%
\Phi}}{(n-1)(n-3)}}\left[ 1-\sqrt{1+\frac{e^{\frac{16\alpha \bar{\Phi}}{%
(n-1)(n-3)}}\bar{\mathcal{F}}}{2\beta ^{2}}}\right] .  \label{LFPBI}
\end{equation}%
We should note that the mentioned Lagrangian is the same as that
in Ref. \cite{sh551}. It is worthwhile to mention that although
the Lagrangian presented in Ref.
\cite{BIdilatonIncorrect1,BIdilatonIncorrect2,BIdilatonIncorrect3}
has correct Maxwell limit for $\beta \longrightarrow \infty $, is
not consistent with conformal transformation and Jordan frame
(this point was indicated in Ref. \cite{sh551}). As it was
presented in Ref. \cite{sh551}, the correct field equations in
which arisen from the variation principle of action can be written
as
\begin{eqnarray}
\bar{R}_{\mu \nu } &=&\frac{4}{n-1}\left( \bar{\nabla}_{\mu }\bar{\Phi}\bar{%
\nabla}_{\nu }\bar{\Phi}+\frac{1}{4}\bar{V}(\bar{\Phi})\bar{g}_{\mu \nu
}\right) -\frac{1}{n-1}\bar{L}(\overline{F},\overline{\Phi })\overline{g}%
_{\mu \nu }  \nonumber \\
&&+\frac{2 e^{-\frac{4\alpha \overline{\Phi }}{n-1}}}{\sqrt{1+\overline{Y}}}%
\left( \overline{F}_{\mu \eta }\overline{F}_{\nu }^{\eta }-\frac{\overline{F%
}}{n-1}\overline{g}_{\mu \nu }\right) ,  \label{Eq1BI}
\end{eqnarray}

\begin{equation}
\bar{\nabla}^{2}\bar{\Phi}=\frac{n-1}{8}\frac{\partial \bar{V}(\bar{\Phi})}{%
\partial \bar{\Phi}}+\frac{\alpha }{2(n-3)}\left( (n+1)\bar{L}(\overline{F},%
\overline{\Phi })+\frac{4e^{-\frac{4\alpha \overline{\Phi }}{n-1}}\overline{F%
}}{\sqrt{1+\overline{Y}}}\right) ,  \label{Eq2BI}
\end{equation}

\begin{equation}
\overline{\nabla }_{\mu }\left( \frac{e^{-\frac{4\alpha \overline{\Phi }}{n-1%
}}}{\sqrt{1+\overline{Y}}}\overline{F}^{\mu \nu }\right) =0,  \label{Eq3BI}
\end{equation}%
where%
\begin{equation}
\overline{Y}=\frac{e^{\frac{16\alpha \overline{\Phi }}{(n-1)(n-3)}}\overline{%
F}}{2\beta ^{2}}.\nonumber
\end{equation}

\subsection{Power Maxwell invariant (PMI) source:}

One of the special classes of nonlinear electrodynamics is Power
Maxwell Invariant (PMI) theory
\cite{HendiBlackString,PMIpaper1,PMIpaper2,PMIpaper3,PMIpaper4,PMIpaper5,PMIpaper6,PMIpaper7,PMIpaper8}.
The PMI theory has an interesting result which distinguishes this
nonlinear theory from others; this theory enjoys conformal
invariancy when the power of Maxwell invariant is a quarter of
spacetime dimensions ($power = dimensions/4$).  It is worth
mentioning that the idea is to take advantages of the conformal
symmetry to construct the analogues of the $4$ dimensional
Reissner-Nordstr\"{o}m solutions with an inverse square law for
the electric field of the point-like charges in arbitrary
dimensions.

Now, we take into account the Lagrangian of nonlinear PMI model
with the following explicit form
\cite{HendiBlackString,PMIpaper1,PMIpaper2,PMIpaper3,PMIpaper4,PMIpaper5,PMIpaper6,PMIpaper7,PMIpaper8}
\begin{equation}
\mathcal{L}(\mathcal{F})=\left( -\mathcal{F}\right) ^{s},  \label{LPMI}
\end{equation}%
and thus one finds $\bar{L}(\bar{Y})=\left( \bar{Y}\right) ^{s}$ with $A=1$
and $B=-1$. Taking into account Eqs. (\ref{LFP2}) and (\ref{Y2}), we find%
\begin{equation}
\bar{L}(\bar{\mathcal{F}},\bar{\Phi})=e^{-\frac{4(n+1)\alpha \bar{\Phi}}{%
(n-1)(n-3)}}\left( -e^{\frac{16\alpha \bar{\Phi}}{(n-1)(n-3)}}\bar{\mathcal{F%
}}\right) ^{s},  \label{LFPPMI}
\end{equation}%
which is coupled Lagrangian of dilaton-PMI theory. One can apply the
variation principle to the corresponding action to obtain the equation of
motions with the following explicit forms
\begin{eqnarray}
\overline{R}_{\mu \nu } &=&\frac{4}{n-1}\left( \bar{\nabla}_{\mu }\bar{\Phi}\bar{%
\nabla}_{\nu
}\bar{\Phi}+\frac{1}{4}\bar{V}(\bar{\Phi})\overline{g}_{\mu \nu
}\right) -\frac{\overline{Y}^{s}}{n-1}e^{-\frac{4(n+1)\alpha \overline{\Phi }%
}{(n-1)(n-3)}}\overline{g}_{\mu \nu }  \nonumber \\
&&+2s\overline{Y}^{s-1}e^{-\frac{4\alpha \overline{\Phi }}{n-1}}\left(
\overline{F}_{\mu \eta }\overline{F}_{\nu }^{\eta }-\frac{\overline{F}}{n-1}%
\overline{g}_{\mu \nu }\right) ,  \label{Eq1PMI}
\end{eqnarray}

\begin{equation}
\bar{\nabla}^{2}\bar{\Phi}=\frac{n-1}{8}\frac{\partial \bar{V}(\bar{\Phi})}{%
\partial \bar{\Phi}}+\frac{\alpha (n-4s+1)}{2(n-3)}e^{-\frac{ 4(n+1) \alpha
\overline{\Phi }}{(n-1)(n-3)}}\overline{Y}^{s},  \label{Eq2PMI}
\end{equation}

\begin{equation}
\overline{\nabla }_{\mu }\left( e^{\frac{\left[ 16s-4(n+1)\right] \alpha
\overline{\Phi }}{(n-1)(n-3)}}(-\overline{F})^{s-1}\overline{F}^{\mu \nu
}\right) =0,  \label{Eq3PMI}
\end{equation}%
where%
\begin{equation}
\overline{Y}=-e^{\frac{16\alpha \overline{\Phi
}}{(n-1)(n-3)}}\overline{F}.\nonumber
\end{equation}

\subsection{Logarithmic theory}

Soleng theory is one of the nonlinear electrodynamics with
logarithmic form \cite{Soleng}. It is also remarkable that this
theory is one of the BI-type theories and enjoys most of the BI
properties such as the absence of shock waves, birefringence
phenomena and respecting an electric-magnetic duality
\cite{NLEDproperties1,NLEDproperties2,NLEDproperties3}. In
addition, like BI electrodynamics, the self-energy of the
point-like charges is finite in this theory. The explicit form of
Soleng Lagrangian is
\begin{equation}
\mathcal{L}(\mathcal{F})=-8\beta ^{2}\ln \left(
1+\frac{\mathcal{F}}{8\beta ^{2}}\right),  \label{LLOG}
\end{equation}
where we set $\bar{L}(\bar{Y})=\ln \left( 1+\bar{Y}\right)$ with
$A=-8\beta ^{2}$ and $B=\frac{1}{8\beta ^{2}}$. Applying the
mentioned function with Eqs. (\ref{LFP2}) and (\ref{Y2}), we
obtain
\begin{equation}
\bar{L}(\bar{\mathcal{F}},\bar{\Phi})=-8\beta ^{2}e^{-\frac{4(n+1)\alpha
\bar{\Phi}}{(n-1)(n-3)}}\ln \left( 1+\frac{e^{\frac{16\alpha \bar{\Phi}}{%
(n-1)(n-3)}}\bar{\mathcal{F}}}{8\beta ^{2}}\right) ,  \label{LFPLOG}
\end{equation}%
which is coupled Lagrangian of dilaton-Soleng theory. Varying the
corresponding action leads to the following field equations%
\begin{eqnarray}
\bar{R}_{\mu \nu } &=&\frac{4}{n-1}\left( \bar{\nabla}_{\mu }\bar{\Phi}\bar{%
\nabla}_{\nu }\bar{\Phi}+\frac{1}{4}\bar{V}(\bar{\Phi})\bar{g}_{\mu \nu
}\right) +\frac{8\beta ^{2}\ln (1+\overline{Y})}{n-1}e^{-\frac{4(n+1)\alpha
\overline{\Phi }}{(n-1)(n-3)}}\overline{g}_{\mu \nu }  \nonumber \\
&&+\frac{2e^{-4\alpha \overline{\Phi }/(n-1)}}{1+\overline{Y}}\left(
\overline{F}_{\mu \eta }\overline{F}_{\nu }^{\eta }-\frac{\overline{F}}{n-1}%
\overline{g}_{\mu \nu }\right) ,  \label{Eq1LOG}
\end{eqnarray}

\begin{equation}
\bar{\nabla}^{2}\bar{\Phi}=\frac{n-1}{8}\frac{\partial \bar{V}(\bar{\Phi})}{%
\partial \bar{\Phi}}-\frac{\alpha }{2(n-3)}\left[ (n+1)\overline{L}(%
\overline{F},\overline{\Phi })+\frac{4e^{-\frac{4\alpha \overline{\Phi }}{n-1%
}}}{1+\overline{Y}}\overline{F}\right] ,  \label{Eq2LOG}
\end{equation}

\begin{equation}
\overline{\nabla }_{\mu }\left( \frac{e^{-\frac{4\alpha \overline{\Phi }}{n-1%
}}}{1+\overline{Y}}\overline{F}^{\mu \nu }\right) =0,  \label{Eq3LOG}
\end{equation}%
where%
\begin{equation}
\overline{Y}=\frac{e^{\frac{16\alpha \overline{\Phi }}{(n-1)(n-3)}}\overline{%
F}}{8\beta ^{2}}. \nonumber
\end{equation}

\subsection{Exponential theory}

In addition to BI and logarithmic types for the nonlinear abelian
gauge field, very recently one of the present authors proposed an
exponential form of nonlinear electrodynamics
\cite{HendiJHEP1,HendiJHEP2}. However, unlike BI and Soleng
theories, the Exponential form of electrodynamics can not deal
with the problem of self-energy completely, its electric field
singularity is much weaker than in the Einstein-Maxwell theory.

Here, we consider exponential form of nonlinear electrodynamics
\cite{HendiJHEP1,HendiJHEP2} and apply our prescription to obtain
its correct coupling with dilaton field in the Einstein frame
\begin{equation}
\mathcal{L}(\mathcal{F})=\beta ^{2}\left( \exp (-\frac{\mathcal{F}}{\beta
^{2}})-1\right) .  \label{LEXP}
\end{equation}%
Regarding this model, we find $\bar{L}(\bar{Y})=\exp (\bar{Y})-1$\ with $%
A=\beta ^{2}$\ and $B=-\frac{1}{\beta ^{2}}$. Taking into account Eqs. (\ref%
{LFP2}) and (\ref{Y2}), one can find
\begin{equation}
\bar{L}(\bar{\mathcal{F}},\bar{\Phi})=\beta ^{2}e^{-\frac{4(n+1)\alpha \bar{%
\Phi}}{(n-1)(n-3)}}\left[ \exp \left( -\frac{e^{\frac{16\alpha \bar{\Phi}}{%
(n-1)(n-3)}}\bar{\mathcal{F}}}{\beta ^{2}}\right) -1\right] ,
\label{LFPhendi}
\end{equation}%
which is coupled Lagrangian of dilaton gravity with the nonlinear
electrodynamics proposed by Hendi \cite{HendiJHEP1,HendiJHEP2}. It
is worthwhile to mention that, although in the special case
($\alpha \longrightarrow 0$) the Lagrangian of Ref.
\cite{SheykhiHajkhalili} reduces to the correspondence Lagrangian
of \cite{HendiJHEP1,HendiJHEP2}, it is not consistent with BD
theory. Whereas proposed Lagrangian of Eq. (\ref{LFPhendi}) has
both requirement conditions, suitable special limits and
consistency with BD theory. Using the variation principle, one
finds
\begin{eqnarray}
\bar{R}_{\mu \nu } &=&\frac{4}{n-1}\left( \bar{\nabla}_{\mu }\bar{\Phi}\bar{%
\nabla}_{\nu }\bar{\Phi}+\frac{1}{4}\bar{V}(\bar{\Phi})\bar{g}_{\mu \nu
}\right) -\frac{\beta ^{2}\left[ \exp (\overline{Y})-1\right] }{n-1}e^{-%
\frac{4(n+1)\alpha \overline{\Phi }}{(n-1)(n-3)}}\overline{g}_{\mu \nu }
\nonumber \\
&&+2e^{-4\alpha \overline{\Phi }/(n-1)}\exp (\overline{Y})\left( \overline{F%
}_{\mu \eta }\overline{F}_{\nu }^{\eta }-\frac{\overline{F}}{n-1}\overline{g}%
_{\mu \nu }\right) ,  \label{Eq1EXP}
\end{eqnarray}

\begin{equation}
\bar{\nabla}^{2}\bar{\Phi}=\frac{n-1}{8}\frac{\partial \bar{V}(\bar{\Phi})}{%
\partial \bar{\Phi}}+\frac{\alpha }{2(n-3)}\left[ (n+1)\overline{L}(
\overline{F},\overline{\Phi })+4e^{-\frac{4\alpha \overline{\Phi }
}{n-1}}\exp (\overline{Y})\overline{F}\right] ,  \label{Eq2EXP}
\end{equation}

\begin{equation}
\overline{\nabla }_{\mu }\left( e^{-\frac{4\alpha \overline{\Phi }}{n-1}%
}\exp (\overline{Y})\overline{F}^{\mu \nu }\right) =0,
\label{Eq3EXP}
\end{equation}%
where%
\begin{equation}
\overline{Y}=\frac{-e^{\frac{16\alpha \overline{\Phi }}{(n-1)(n-3)}}\overline{%
F}}{\beta ^{2}}.\nonumber
\end{equation}

\section{BD versus dilaton solutions}

Having the suitable field equations, we are in a position to take into
account the field equations to obtain suitable metric for the spacetime.
Solving the field equations of dilaton gravity in Einstein frame is not hard
and one can obtain consistent functions for $\bar{g}_{\mu \nu }$, $\bar{F}%
_{\mu \nu }$ and $\bar{\Phi}$, depending the initial conditions of
spacetime, such as spherical or cylindrical symmetry.

By assuming the $\left( \bar{g}_{\mu \nu },\bar{F}_{\mu \nu },\bar{\Phi}%
\right) $ as solutions of Eqs. (\ref{fieldc1})-(\ref{fieldc3}) with
potential $\bar{V}\left( \bar{\Phi}\right) $ and comparing Eqs. (\ref%
{field01})-(\ref{field03}) with Eqs. (\ref{fieldc1})-(\ref{fieldc3}) we find
the solutions of Eqs. (\ref{field01})-(\ref{field03}) with potential $V(\Phi
)$ can be written as
\begin{equation}
\left[ g_{\mu \nu },F_{\mu \nu },\Phi \right] =\left[ \exp \left( -\frac{%
8\alpha \bar{\Phi}}{\left( n-1\right) (n-3)}\right) \bar{g}_{\mu \nu },\bar{F%
}_{\mu \nu },\exp \left( \frac{4\alpha \bar{\Phi}}{n-3}\right) \right] .
\label{sol}
\end{equation}

\section{Implementation of second case to an energy dependent Ricci-flat spacetime}

Now, we are in a position to apply previous results to a typical
spacetime. In order to enrich the physical properties of the
solutions, we consider $(n+1)-$dimensional energy dependent
spacetime with flat boundary. In other words, we want to obtain
black hole solutions of BI-Dilaton gravity's rainbow and their
BD-BI counterpart. We start with the following metric
\begin{equation}
ds^{2}=-\frac{N(r)}{f^{2}(\varepsilon )}dt^{2}+\frac{1}{g^{2}(\varepsilon )}%
\left( \frac{dr^{2}}{N(r)}+{r^{2}R(r)}^{2}d\Omega _{n-1}^{2}\right) ,
\label{Metric}
\end{equation}%
where $r^{2}d\Omega _{n-1}^{2}=r^{2}\sum_{i=1}^{n-1}d\theta _{i}^{2}$ is an $%
(n-1)$-dimensional Euclidean space, $N(r)$\ is the metric functions and also
$f(\varepsilon )$ and $g(\varepsilon )$\ are rainbow functions. We should
introduce a suitable potential to solve the field equations, (\ref{Eq1BI})
and (\ref{Eq3BI}), simultaneously. Taking into account Eq. (\ref{Eq3BI})
with metric (\ref{Metric}), we find the following differential equation%
\begin{equation}
\left[ \frac{4\alpha \overline{\Phi }^{\prime }}{(n-1)}-\left( \frac{%
R^{\prime }}{R}+\frac{1}{r}\right) \right] F_{tr}^{2}e^{\frac{16\alpha
\overline{\Phi }}{(n-1)(n-3)}}-n\beta ^{2}\left[ \frac{4\alpha \overline{%
\Phi }^{\prime }}{(n-1)^{2}}-(n-1)^{2}\frac{R^{\prime }}{R}-\left( \frac{1}{r%
}+\frac{1}{(n-1)R}\frac{F_{tr}^{\prime }}{F_{tr}}\right) \right]
=0, \nonumber
\end{equation}%
where its solution is for the electric field is%
\begin{equation}
F_{tr}(r)=E(r)=\frac{q\exp \left( \frac{4\alpha \overline{\Phi }(r)}{n-1}%
\right) }{[rR(r)]^{n-1}\sqrt{1+\frac{f^{2}(\varepsilon
)g^{2}(\varepsilon )q^{2}\exp \left( \frac{8\alpha \overline{\Phi
}(r)}{n-3}\right) }{\beta ^{2}[rR(r)]^{2(n-1)}}}}.\nonumber
\end{equation}

Regarding a suitable ansatz $R(r)=\exp \left( \frac{2\alpha \overline{\Phi }%
}{n-1}\right) $, one finds the difference between $tt$ and $rr$
components of Eq. (\ref{Eq1BI}) as
\begin{equation}
\overline{\Phi }^{\prime \prime }+2\overline{\Phi }^{\prime }\left( \frac{%
(1+\alpha ^{2})}{\alpha (n-1)}\overline{\Phi }^{\prime }+\frac{1}{r}\right)
=0  \label{PhiEq}
\end{equation}%
with the following solution for the dilaton field%
\begin{equation}
\overline{\Phi }(r)=\frac{(n-1)\gamma }{2\alpha }\ln \left( \frac{b}{r}%
\right) .  \label{Phi}
\end{equation}

Now, we should find the metric function $N(r)$. To do so, we
should set a suitable potential, $\overline{V}(\overline{\Phi })$.
For dilatonic Maxwell Lagrangian, suitable potential is
$\overline{V}(\overline{\Phi })=2\Lambda \exp \left( \frac{4\alpha
\overline{\Phi }(r)}{n-1}\right) $, in which motivates us to
consider the following potential
\begin{equation}
\overline{V}(\overline{\Phi })=2\Lambda e^{\left( \frac{4\alpha \overline{%
\Phi }(r)}{n-1}\right) }+\frac{W(r)}{\beta ^{2}},  \label{Pot}
\end{equation}%
where for the Maxwell limit ($\beta \rightarrow \infty $), second term
vanishes. In order to obtain metric functions and unknown $W(r)$, we regard $%
\theta _{i}\theta _{i}$ component of Eq. (\ref{Eq1BI}) with Eq. (\ref{Eq2BI}%
) to obtain, respectively,
\begin{eqnarray}
&&\left( \frac{R^{\prime }}{R}-\frac{1}{r}\right) \frac{N^{\prime }}{N}-%
\frac{R^{\prime \prime }}{R}-\frac{(n-2)R^{\prime \prime }}{R^{2}}-\frac{%
2(n-1)R^{\prime
}}{rR}-\frac{(n-2)}{r^{2}}-\frac{\overline{V}(\overline{\Phi
})}{\left( n-1\right) Ng^{2}(\varepsilon )}+ \nonumber \\
&&\frac{4\beta ^{2}e^{-\frac{4(n+1)\alpha \overline{\Phi }}{(n-1)(n-3)}}}{%
\left( n-1\right) Ng^{2}(\varepsilon )}\left[ 1-\frac{1}{\sqrt{1-\frac{e^{%
\frac{16\alpha \overline{\Phi }}{(n-1)(n-3)}}f^{2}(\varepsilon
)g^{2}(\varepsilon )F_{tr}^{2}}{\beta ^{2}}}}\right]  =0 \\
\label{ththEq} \nonumber \\ \nonumber \\
&&\overline{\Phi }^{\prime \prime }+\overline{\Phi }^{\prime }\left[ (n-1)(%
\frac{1}{r}+\frac{R^{\prime }}{R})+\frac{N^{\prime }}{N}\right] +\frac{%
4\alpha qf(\varepsilon )F_{tr}}{(n-3)(rR)^{(n-1)}}-\frac{n-1}{%
8Ng^{2}(\varepsilon )}\left( \frac{d}{d\overline{\Phi }}\overline{V}(%
\overline{\Phi })\right) - \nonumber \\
&&\frac{2(n+1)\alpha \beta ^{2}e^{\left( -\frac{4\alpha \overline{\Phi }(n+1)}{%
(n-1)(n-3)}\right) }}{\left( n-3\right) Ng^{2}(\varepsilon )}\left[ 1-\frac{1%
}{\sqrt{1-\frac{e^{\frac{16\alpha \overline{\Phi }}{(n-1)(n-3)}%
}f^{2}(\varepsilon )g^{2}(\varepsilon )F_{tr}^{2}}{\beta
^{2}}}}\right]  =0. \label{Del2PhiEq}
\end{eqnarray}

It is a matter of calculation to show that the following functions
satisfy the field equations
\begin{equation}
W(r)=\frac{4q(n-1)\beta ^{2}f^{2}(\varepsilon )g^{2}(\varepsilon )R}{\left(
\alpha ^{2}+1\right) r^{\gamma }b^{n\gamma }}\int \frac{F_{tr}}{%
r^{n(1-\gamma )-\gamma }}dr+\frac{4\beta
^{4}}{R^{\frac{2(n+1)}{n-3}}}\left(
1-\frac{F_{tr}R^{n-3}r^{n-1}}{q}\right) -\frac{4q\beta
^{2}f^{2}(\varepsilon )g^{2}(\varepsilon )F_{tr}}{r^{n-1}}\left(
\frac{b}{r}\right) ^{-(n-1)\gamma}
\end{equation}
\begin{eqnarray}
N(r) &=&\left( \frac{(1+\alpha ^{2})^{2}r^{2}}{(n-1)}\right) \frac{2\Lambda
\left( \frac{r}{b}\right) ^{-2\gamma }}{g^{2}(\varepsilon )(\alpha ^{2}-n)}-%
\frac{m}{r^{(n-1)(1-\gamma )-1}}- \\
&&\frac{4(1+\alpha ^{2})^{2}q^{2}(\frac{r}{b})^{2\gamma
(n-2)}}{(n-\alpha ^{2})r^{2(n-2)}}\left(
\frac{1}{2(n-1)}F_{1}(\eta )-\frac{1}{\alpha ^{2}+n-2}F_{2}(\eta
)\right) ,
\end{eqnarray}%
where%
\begin{eqnarray}
R(r) &=&\left( \frac{b}{r}\right) ^{\gamma }, \nonumber \\
F_{tr} &=&\frac{q}{r^{n-1}}\frac{\left( \frac{b}{r}\right)
^{\frac{2-\gamma ^{2}(n-1)}{\gamma
}}}{\sqrt{1+\frac{f^{2}(\varepsilon )g^{2}(\varepsilon
)q^{2}}{\beta ^{2}r^{2n-2}\left( \frac{b}{r}\right)
^{\frac{2(n-1)[\gamma
^{2}(n-3)-2]}{\gamma (n-3)}}}}}, \nonumber \\
F _{1}(\eta) &=&_{2}F_{1}\left( \left[ \frac{1}{2},\frac{(n-3)\Upsilon }{%
\alpha ^{2}+n-2}\right] ,\left[ 1+\frac{(n-3)\Upsilon }{\alpha ^{2}+n-2}%
\right] ,-\eta \right) , \nonumber \\
F _{2}(\eta ) &=&_{2}F_{1}\left( \left[ \frac{1}{2},\frac{(n-3)\Upsilon }{%
2(n-1)}\right] ,\left[ 1+\frac{(n-3)\Upsilon }{2(n-1)}\right]
,-\eta \right)
, \nonumber \\
\eta &=&\frac{g^{2}(\varepsilon )f^{2}(\varepsilon )q^{2}(\frac{r}{b}%
)^{2\gamma (n-1)(n-5)/(n-3)}}{\beta ^{2}r^{2(n-1)}}, \nonumber \\
\Upsilon &=&\frac{\alpha ^{2}+n-2}{2\alpha ^{2}+n-3}.\nonumber
\end{eqnarray}%
It is notable that $b$ and $m$ are integration constants, $\gamma
=\alpha ^{2}/\left( 1+\alpha ^{2}\right) $ and
$_{2}F_{1}([a,b],[c],z)$ is the hypergeometric function. It is
worthwhile to mention that obtained functions satisfy all field
equations, simultaneously. In addition, one can find that for
large values of the nonlinearity parameter, $\beta $,
\begin{equation}
\left. W(r)\right\vert _{{large \ }\beta }=\frac{g^{4}(\varepsilon
)f^{4}(\varepsilon )q^{4}\alpha ^{4}\left( \frac{b}{r}\right) ^{\frac{%
2\gamma (2n^{2}-11n+11)}{(n-3)}}}{2(n+\alpha ^{2}-3)r^{\frac{10\gamma
(n-1)+2(1-\gamma )(2n^{2}+19n-17)}{(n-3)}}}-\frac{g^{6}(\varepsilon
)f^{6}(\varepsilon )q^{6}\alpha ^{2}\left( \frac{b}{r}\right) ^{\frac{%
2\gamma (3n^{2}-17n+16)}{(n-3)}}}{(3n+4\alpha ^{2}-9)r^{\frac{6(n^{2}-4n+3)}{%
(n-3)}}\beta ^{2}}+O(\frac{1}{\beta ^{4}}), \label{W(r)2}
\end{equation}%
which confirms that the nonvanishing term of Eq. (\ref{W(r)2}) is
the first term for $\beta \rightarrow \infty $. In addition, we
can conclude that the second term of Eq. (\ref{Pot})
($\frac{W(r)}{\beta^2}$) vanishes in this limit ($\beta
\rightarrow \infty $), as we expected. In other words, for $\beta
\rightarrow \infty $, the Maxwell solutions (and also their
corresponding potential ($\overline{V}(\overline{\Phi })=2\Lambda
e^{\left( \frac{4\alpha \overline{\Phi }(r)}{n-1}\right) }$)) will
be recovered.

Now, we can obtain black hole solutions of BD-BI gravity's
rainbow. We consider the following Ricci-flat energy dependent
spacetime
\begin{equation}
ds^{2}=-\frac{A(r)}{f^{2}(\varepsilon )}dt^{2}+\frac{1}{g^{2}(\varepsilon
)B(r)}\left( dr^{2}+r^{2}H(r)^{2}d\Omega _{n-1}^{2}\right) .  \label{Metric2}
\end{equation}

Using the conformal transformation (\ref{sol}), we find that the following
functions satisfy all BD-BI field equations, simultaneously,%
\begin{eqnarray*}
A(r) &=&\left( \frac{b}{r}\right) ^{-4\gamma /(n-3)}N(r), \\
B(r) &=&\left( \frac{b}{r}\right) ^{4\gamma /(n-3)}N(r), \\
H(r) &=&\left( \frac{b}{r}\right) ^{\gamma (n-5)/(n-3)}, \\
\Phi (r) &=&\left( \frac{b}{r}\right) ^{2\gamma (n-1)/(n-3)},
\end{eqnarray*}%
where $V(\Phi )$ can be calculated from Eq. (\ref{PhiPhi}).

\section{Conclusions}

In this paper, we discussed about the conformal relation of scalar
tensor gravity in the Einstein and Jordan frames. The main goal of
this paper was presenting a formal method to obtain the correct
coupling between scalar field and electrodynamics. We introduced a
general method for all linear and nonlinear electrodynamics. We
obtained the correct Lagrangian of dilaton-electrodynamics and
related field equations as well. We also obtained the Lagrangian
of known theory as the cases study and found that our method are
valid for various models of electrodynamics.

We should note that in order to introduce a correct Lagrangian of dilaton
gravity coupled with electrodynamics, we should applied two requirements:
compatibility with BD theory and appropriate special situations. We also
reported that some of dilatonic black hole papers used the inconsistent
Lagrangian and did not care for the BD consistency.

We also, applied proposed method to the case of BI nonlinear electrodynamics
and obtained a suitable potential for consistency of field equations. We
achieved analytical solutions of both dilaton-BI and BD-BI theories in
gravity's rainbow and found the effects of rainbow functions in metric
functions as well as other fields. We left geometrical and thermodynamical
properties of the solutions for an independent work.

In this paper, we restricted ourselves to BD theory. It is worthwhile to
extend BD theory to a general form of scalar tensor gravity in both Einstein
and Jordan frames. We left these issues for the forthcoming work.

\section{acknowledgements}
We would like to thank an anonymous referee for suggesting
important improvements. MST is indebted to M. Kord Zangeneh for
useful discussions. We also thank the Shiraz University Research
Council. This work has been supported financially by the Research
Institute for Astronomy and Astrophysics of Maragha, Iran.

\end{document}